\documentclass{cupconf}

  \checkfont{eurm10}
  \iffontfound
    \IfFileExists{upmath.sty}
      {\typeout{^^JFound AMS Euler Roman fonts on the system,
                   using the 'upmath' package.^^J}%
       \usepackage{upmath}}
      {\typeout{^^JFound AMS Euler Roman fonts on the system, but you
                   dont seem to have the}%
       \typeout{'upmath' package installed. cupconf.cls can take advantage
                 of these fonts,^^Jif you use 'upmath' package.^^J}%
      }
  \else
  \fi

  \checkfont{msam10}
  \iffontfound
    \IfFileExists{amssymb.sty}
      {\typeout{^^JFound AMS Symbol fonts on the system, using the
                'amssymb' package.^^J}%
       \usepackage{amssymb}%
         \let\leq=\leqslant
         \let\geq=\geqslant
      }{}
  \fi

  \IfFileExists{amsbsy.sty}
    {\typeout{^^JFound the 'amsbsy' package on the system, using it.^^J}%
     \usepackage{amsbsy}}
    {\providecommand\boldsymbol[1]{\mbox{\boldmath $##1$}}}

\def\gtorder{\mathrel{\raise.3ex\hbox{$>$}\mkern-14mu
             \lower0.6ex\hbox{$\sim$}}}
\def\ltorder{\mathrel{\raise.3ex\hbox{$<$}\mkern-14mu
             \lower0.6ex\hbox{$\sim$}}}

\def\farcs{\hbox{$.\!\!^{\prime\prime}$}}
\newcommand{\kms}{\mbox{ km~s$^{-1}$}}

\newsavebox{\astrutbox}
\sbox{\astrutbox}{\rule[-5pt]{0pt}{20pt}}

\newcommand\etal{\mbox{\textit{et al.}}}

\title[Dynamical Probes of The Halo Mass Function]
      {Dynamical Probes of The Halo Mass Function}

\author[C.S. Kochanek]{C.S. Kochanek}
\affiliation{Smithsonian Astrophysical Observatory \\
   Harvard-Smithsonian Center for Astrophysics, MS-51 \\ 
   60 Garden Street, Cambridge, MA 02138 \\ 
   ckochanek@cfa.harvard.edu
   }

\input psfig

\begin{document}
\maketitle

\begin{abstract}
We explore the relationship between the mass function of CDM halos and dynamical
probes of the mass function such as the distribution of gravitational lens 
separations and the local velocity function.  The compression of galactic halos by
the cooling baryons, a standard component of modern models, leads to a feature
in the distribution of lens separations near $\Delta\theta=3\farcs0$ or
in the velocity function near $v_c \simeq 400\kms$.  The two probes of the
mass function, lens separations and the local velocity function, are mutually
consistent.  Producing the observed velocity function of galaxies or the
separation distribution using standard adiabatic compression models requires more 
cold baryons, an equivalent cosmological density of $\Omega_{b,{\rm cool}} \simeq 0.02$
compared to a total cosmological $\Omega_b \simeq 0.04$, than are observed in standard 
accountings of the baryonic content of galaxies, $\Omega_{b,gal} \simeq 0.006$,
or our Galaxy, $\Omega_{b,Gal} \ltorder 0.015$. The requirement for a higher cold 
baryon density than is usually assigned to galaxies appears to be generic to models 
which use the standard adiabatic compression models for the transformation of the CDM 
halo by the cooling baryons.  If real, this {\it dynamical baryon discrepancy} suggests 
either that we are neglecting half of the cold baryonic mass in standard galactic models 
(e.g. a MACHO, cold molecular or warm gas component), or that there is a 
problem with standard adiabatic compression models.  
\end{abstract}

\firstsection 
              
\section{Introduction}

In large part due to high resolution N-body simulations, the number density, spatial 
distribution, and properties of dark matter halos are well understood in models based on 
hierarchical clustering (e.g. Jenkins et al. 2000; Sheth \& Torman 1999; Navarro et al. 1996; 
Moore et al. 1998).  The relationship of these halos to astrophysical objects is less
well understood because of the modifications to the halos produced by baryonic physics
and the dependence of our search and measurement techniques on their baryonic properties.
While semi-analytic models of galaxy formation (e.g. Lacey \& Silk 1991, White \& 
Frenk 1991; Cole et al. 1994, Baugh et al. 1996, Kauffman et al. 1993, 1999,
Dalcanton et al. 1997; Somerville \& Primack 1999; Benson et al. 2000; Cole et al. 2000\footnote{We
will collectively refer to the results of semi-analytic models as SA}) model 
these effects with considerable success, the results depend on detailed, parametric 
models for star formation and feedback processes tuned to fit the data.

We would like to have approaches for comparing the properties of halos and astrophysical 
objects which minimize the dependence of the comparison on star formation and luminosity.  
We can generally 
refer to these methods as dynamical probes of dark matter halos because they focus on the
observational properties of the mass distribution rather than of the luminosity
distribution.  This approach is well-developed for massive clusters, where there
are many projects designed to determine the cosmological model and the normalization
of the power-spectrum by comparing the abundance of clusters with the mass function
(see the reviews by Bahcall, Donahue, Rosati \& Tyson in these proceedings).  Here we want to 
focus more on the mass scales of galaxies
and on global comparisons including galaxies, groups and clusters rather than the
simpler case of rich clusters.  The two dynamical probes we can use to relate the
halo mass function to astrophysical objects are the velocity function, the distribution 
of halos in their circular velocity, and the separation distribution of 
gravitational lenses.  

Although we seek tests which avoid any dependence on the luminosity of a halo, we 
cannot avoid the effects of the baryons.  The cooling of the baryons in the lower
mass halos (i.e. galaxies) and the associated adiabatic compression of the 
dark matter (e.g. Blumenthal et al. 1986, SA) significantly alters the density 
distribution of the halo and thus the properties of any dynamical probe of the
halo.  While these effects are included in almost all semi-analytic models, 
we approach the problem from a very different viewpoint.  These models are also
at the center of the controversy over the consistency of the cusped dark matter
profiles predicted by simulations with the observed central rotation curves
of galaxies (e.g. Flores \& Primack 1994, Moore 1994, de Blok et al 2001,
van den Bosch et al. 2000, 2001, Salucci 2001, also Burkert, Sancisi \& Sanders
in these proceedings).  If there is a conflict and it is not due to an problem
in the dark matter density distribution, then it must be due to a problem in
either the adiabatic compression model or the assumed baryon distribution.
The latter possibility is particularly interesting because we know that standard
models of galaxies include only a small fraction of the available baryons.

We start in \S2 with a compressed review of the models we use to determine the
mass function of halos, the adiabatic compression of halos and simple cooling
models for the baryons.  In \S3 we use these models to understand the distribution
of image separations in gravitational lenses.  In \S4 we use the same models
to study the local velocity function of galaxies and clusters.  In \S5 we
make a final check of the consistency of the model by determining the velocity
function from the gravitational lens separation distribution rather than local 
dynamics.  Finally in \S6 we outline a non-parametric approach to understanding the 
relation between the velocity function of galaxies and the mass function of halos.
We discuss the future of dynamical probes in \S7. 

\section{A Very Compressed Theoretical Review}

We use a fixed $\Omega_0=0.3$ $\Lambda$CDM cosmological model with a Hubble constant
of $H_0=67$~km~s$^{-1}$~Mpc$^{-1}$, a baryon density of $\Omega_b=0.04$ and a power
spectrum normalized by the abundance of rich clusters ($\sigma_8=0.9$).  We calculated 
the mass function of the dark matter halos using the Press-Schechter (1974) theory combined 
with the Sheth \& Torman (1999) fit to the results of the Virgo simulations.  Where
needed, we followed Kitayama \& Suto (1996) and Newman \& Davis (2000) in modeling
the distribution of halo formation times using the extended Press-Schechter theory
outlined in Lacey \& Cole (1994).  We neglected the problem of ``halos-in-halos''
(e.g. Peacock \& Smith 2001, Scoccimarro et al. 2001), as it introduces considerable 
complexity for very modest changes in the mass function ($\sim 10\%$, White et al. 2001).
In short, we assume the halo occupancy number is unity for galaxies and neglect accounting
for the galaxies in clusters.

We used the Mo et al. (1998) model for the modifications to the mass distribution created 
by the cooling of the baryons and the adiabatic compression of the dark matter, although
similar approaches are used in all SA studies.  We assume that halos have the NFW
(Navarro et al. 1996) density profile, $\rho \propto 1/x(1+x)^3$, where $x=r/r_s$
is the radius in units of the break radius $r_s$.  Each halo is characterized by
its virial mass $M_{vir}$ and the concentration $c=r_{vir}/r_s$.  The virial mass
is defined by the radius $r_{vir}$ at which the enclosed density exceeds the critical
density by $\Delta_c(z)$. We estimated the concentration by the mean relation
$c\simeq 9 (1+z)^{-1} (M_{vir}/8\times 10^{12}M_\odot)^{-0.14}$ from   
Bullock et al. (2000).  Finally, the
halo is assumed to have angular momentum $J$ specified by its spin parameter 
$\lambda= J|E|^{1/2}/ GM_{vir}^{5/2}$ where the binding energy $|E|$ is computed
using the virial theorem.

We model galaxies as exponential disks with masses of $M_d= m_d M_{vir}$ and 
scale lengths $r_d$.  The disk is assumed to have angular momentum $J_d=j_d J$
and this is used to determine the disk scale length (see Mo et al. 1998). 
Unlike Mo et al. (1998) we added a bulge modeled as a Hernquist (1990) profile
with mass $M_b = m_b M_{vir}$ and an empirically estimated scale length of
$0.045 r_d$ from the photometry of galaxies.  We usually assumed that the
total specific angular momentum of the baryons was the same as that of the 
dark matter but that all the angular momentum is in the disk component 
(i.e. $j_d=m_d+m_b$ and $j_b=0$).  We use the standard Blumenthal et al. (1986)
model for the adiabatic compression of the dark matter by the cooled baryons.

Finally, we used a simplified version of the Cole et al. (2000) cooling model 
from their semi-analytic models.  We are interested in cooling because it plays
an important role in determining the boundary between galaxies (where the gas
can cool and form stars) and groups/clusters (where it remains hot).  The
Cole et al. (2000) model provides an estimate of the cooling time as a function
of radius in a halo of a given virial temperature (i.e. mass), 
$\tau_{cool}(M,r)$, and we determine the cooled baryonic mass fraction 
$f_{cool}(M,z)$ at redshift $z$ by the mass fraction inside the radius
where the cooling time equals the current age, 
$t(z)-t_{form}(M,z)=\tau_{cool}(M,r_{cool})$.  If the global baryon fraction
is $(m_d+m_b)_0$ then we model the halo by an adiabatic compression model
of cold baryon mass fraction $m_d+m_b = (m_d+m_b)_0 f_{cool}(M,z)$.  Assuming halos
start as fair samples of the universe, the global baryon fraction is
$(m_d+m_b)_0=\Omega_b/\Omega_0 = 0.13$.  However, the final cold baryon 
fraction can be smaller, $\Omega_{b,{\rm cool}} \leq \Omega_b$, because star
formation and feedback can reheat the baryons which initially cooled (see
SA).  

For our scalings, the peak rotation velocity of an NFW halo of mass $M$ is
\begin{equation}
    v_{mod,0}(M) = 186 (M/10^{12} M_\odot)^{0.30} \kms
\end{equation}
which rises to
\begin{equation}
        v_{mod}(M,m_d,\lambda) = v_{mod,0}(M)
    \left[ 1 + { 314 m_d^2 \Lambda^{-0.98} \over 1 + 42.3 m_d \Lambda^{0.30}} \right]
\end{equation}
for a halo with cold baryon fraction $m_d$ (disk only, $m_b=0$) and an effective spin
parameter of  $\Lambda = (j_d/m_d)(\lambda/\bar{\lambda})$ normalized by a mean 
spin parameter of $\bar{\lambda}=0.05$ and valid for $0\leq m_d \leq 0.15$ and 
$0.02 \leq \lambda \leq 0.1$ (see Mo et al. 1998 for the effects of varying 
$c$).

\begin{figure}
\centerline{\psfig{figure=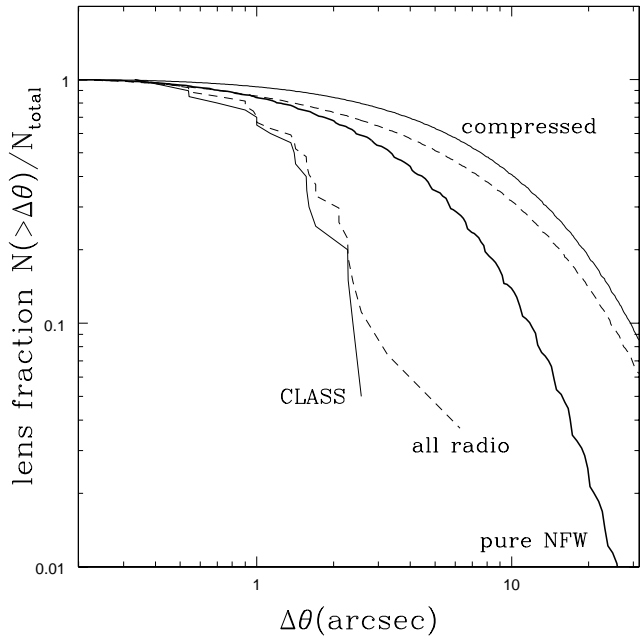,width=3.5in}}
\caption{
Predicted separation distributions without a cooling scale.
The curves show the fraction of lenses with separations exceeding
$\Delta\theta$, $N(>\Delta\theta)/N_{\rm total}$.  The observed distributions
are shown by the curves labeled CLASS (for the CLASS survey lenses) and
all radio (for all radio-selected lenses).
The heavy solid line shows the distribution
predicted by pure NFW models while the light solid (dashed) lines shows the
distributions predicted by the adiabatic compression models with no bulge
(a 10\% baryonic mass fraction bulge). }
\end{figure}

\begin{figure}
\centerline{\psfig{figure=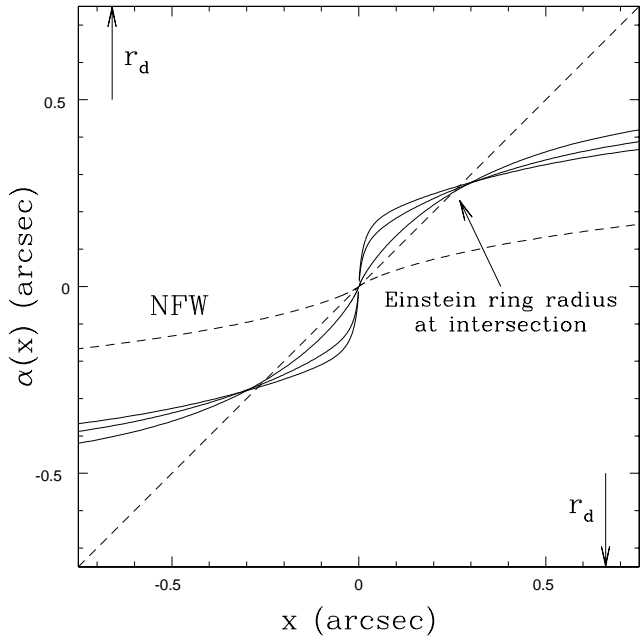,width=3.5in}}
\caption{
The bending angles, $\alpha(x)$, produced by a $10^{12}M_\odot$ halo at
$z_l=0.5$ with concentration $c=8$ for a source at $z_s=2.0$.  The
dashed curve shows the bend angle for the initial sub-critical NFW halo.  The
solid curves show the bend angles found after the baryons cool
assuming a baryonic mass fraction of $m_d+m_b=0.05$, a spin parameter
$\lambda=0.04$, and that the disk contains all the initial baryonic angular
momentum, $j_d=0.05$.  The three solid curves are for bulge-to-disk mass
ratios $m_b/m_d=0$ (shallowest rise), $0.1$ and $0.2$ (steepest rise)
respectively.
The tangential critical line of the lens (the Einstein ring) is located at
the point where the (dashed) $45^\circ$ line intersects the bending angle,
and the radial critical line is located where a $45^\circ$ line is tangent
to the bending angle.
An arrow points to the location of the tangential critical line of the
adiabatically compressed models.  The arrows on either side of the figure
indicate the (angular) disk scale length $r_d$ at the model redshift.}
\end{figure}

\begin{figure}
\centerline{\psfig{figure=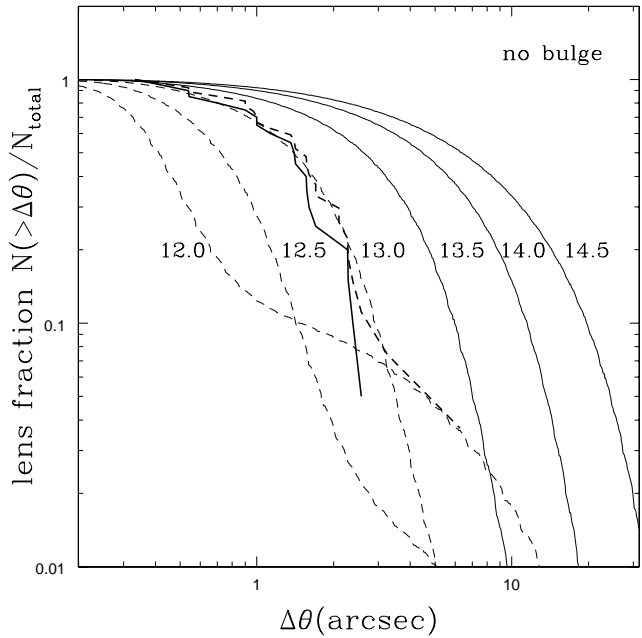,width=3.5in}}
\caption{
Predicted separation distributions with a cooling scale for models
without a bulge and cold baryon fraction $m_d+m_b=0.05$ in halos
below a cooling mass scale $M_c$.  The curves are labeled by
$\log M_c/M_\odot$.  The heavy solid (dashed) curve shows the observed distribution
of the CLASS (radio-selected) lenses. }
\end{figure}

\begin{figure}
\centerline{\psfig{figure=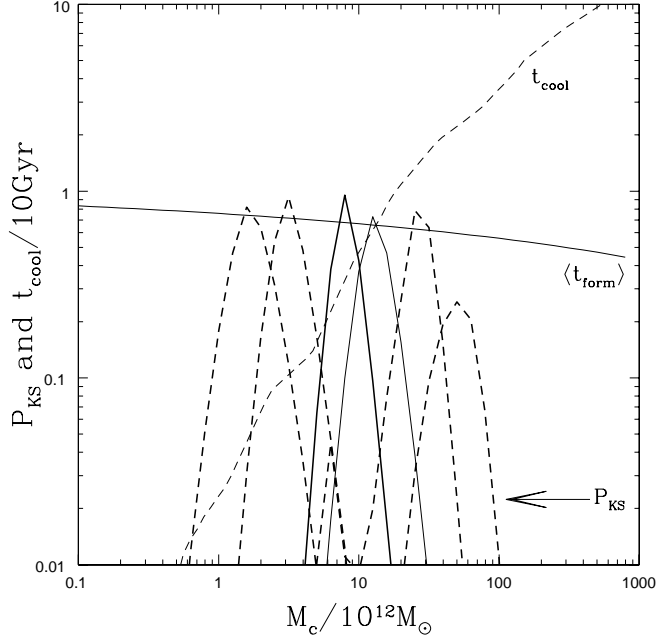,width=3.5in}}
\caption{ The Kolmogorov-Smirnov probability, $P_{KS}$, of fitting the
  observed separation distribution of CLASS lenses as a function of the
  cooling mass scale $M_c$. The heavy (light) solid curves indicated by
  the arrow show the K-S probability for models with $m_b+m_d=0.05$
  without (with) a $m_b/m_d=0.10$ bulge.  The heavy dashed curves show
  the K-S probabilities for models with lower ($m_b+m_d=0.01$ and $0.02$)
  or higher ($m_b+m_d=0.10$ and $0.20$) baryon fractions where
  the optimal cooling mass decreases as the baryon fraction rises.  
  The light dashed curves show the cooling time in units of $10$~Gyr for 
  the radii enclosing 50\% of the baryonic mass for the standard model.
  The light solid line shows the time since the average formation epoch
  ($\langle t_{\rm form}\rangle$) in units of $10$~Gyr assuming $h=0.67$.}
\end{figure}

\begin{figure}
\centerline{\psfig{figure=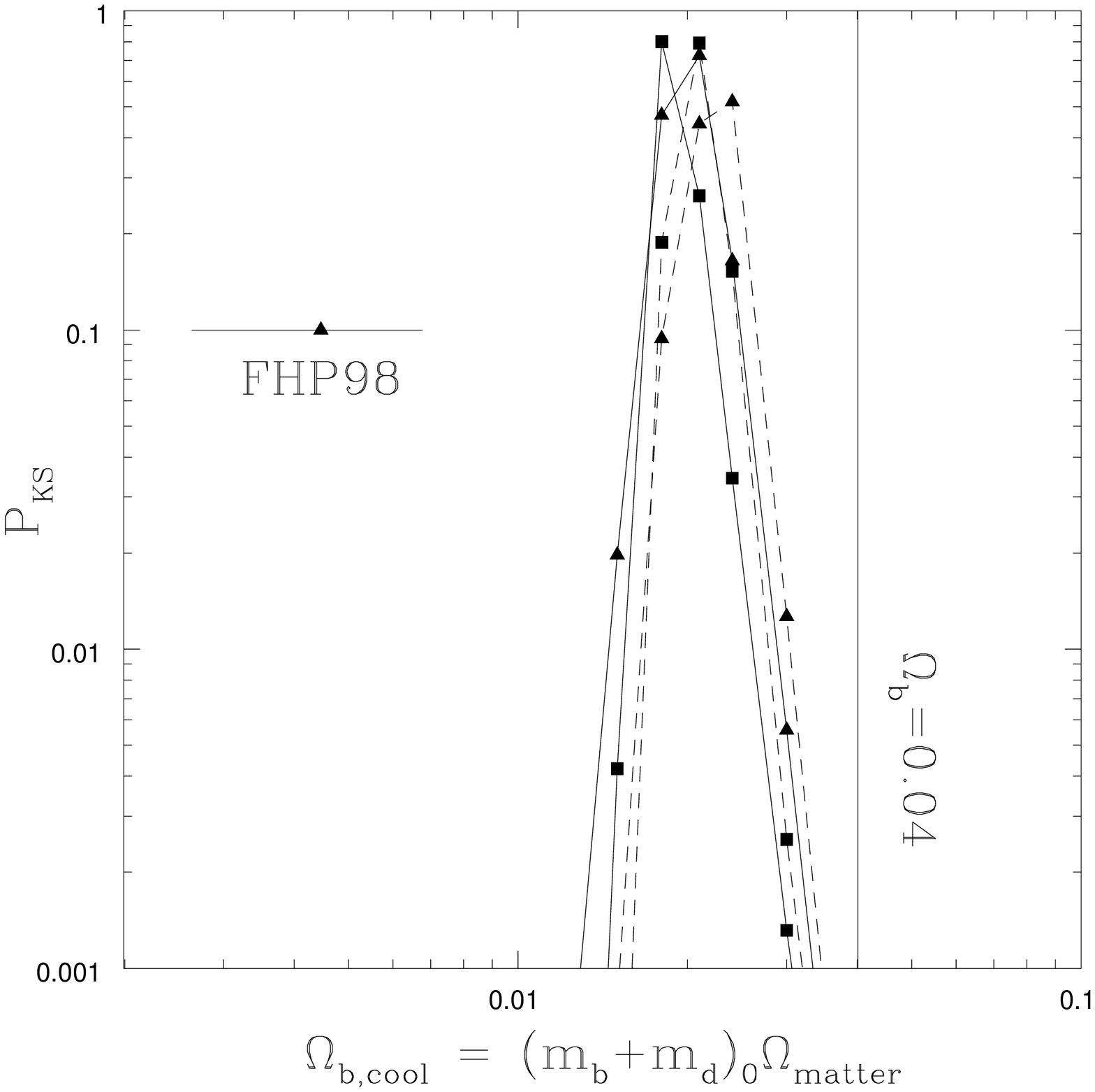,width=3.5in}}
\caption{ Kolmogorov-Smirnov test probability of fitting the separation
distribution of CLASS lenses as a function of $\Omega_{b,{\rm cool}}$.
The squares (triangles) indicate models with no bulge (with a $m_b/m_d=0.1$
bulge), and the solid (dashed) lines correspond to fitting the CLASS lenses
(all radio lenses).  The point with horizontal error bar is the estimate
by Fukugita, Hogan \& Peebles~(\protect\cite{Fuk98}) for the cold baryon
(stars, remnants, cold gas) content of galaxies.  The vertical line marks
the total baryon content in the concordance model. }
\end{figure}

\section{Why Don't Cluster Lenses Exist?}

One of the most striking features of surveys for gravitational lenses is that
cluster lenses do not exist.  This statement may seem peculiar given the
enormous attention devoted to lensing by rich clusters (e.g. Tyson in these
proceedings), but it is simply another facet of the fact that rich clusters 
loom large in our imaginations despite being exponentially rare and 
containing a negligible galaxy or mass fraction.  The known rich cluster lenses were all found by 
first finding a rich cluster and then searching for lensed sources behind them.
  
Only surveys which examine sources to see if they are lensed probe the halo
mass function, because there is a mapping between the image separation 
distribution $dn/d\Delta\theta$ and the halo mass function $dn/dM$.  For
example, the CLASS survey (e.g. Browne \& Myers 2000, Philips et al. 2000) 
for lensed flat-spectrum radio sources has a nearly uniform selection function
from $0\farcs3 \ltorder \Delta\theta \ltorder 15\farcs0$ and has found 18
lenses all with separations $\Delta\theta <3\farcs0$.  If we consider all
27 radio-selected lenses, there are two wider separation lenses (MG2016+112
and Q0957+561) reaching to $\Delta\theta \ltorder 6\farcs0$.  Despite having the
sensitivity to wide separations needed to find cluster lenses, the 
distribution is overwhelmingly dominated by galaxy lenses 
(average separations of 1\farcs5) with a few lenses due to groups and 
poor clusters on larger scales (see Fig. 1).

All calculations of the separation distribution of lenses 
combining the halo mass function with a model density distribution for the
halos catastrophically fail to explain the observed distribution of image
separations (e.g. Narayan \& White 1988, Kochanek 1995, Wambsganss et al.
1995, 1998, Maoz et al. 1997, Keeton 1998, Mortlock \& Webster 2000,
Li \& Ostriker 2000, Keeton \& Madau 2000, Wyithe et al. 2000).  Models 
normalized by the abundance of rich clusters correctly find that rich
cluster lenses are rare, but then grossly under predict the number of 
galaxy-scale lenses (see Fig. 1).  
A purely phenomenological approach based on the local 
properties of galaxies, by contrast, predicts the observed properties of 
the lenses well (e.g. Kochanek 1996, Keeton et al. 1998).  These
models have modest difficulty explaining the largest lenses found in 
systematic surveys ($\Delta\theta\simeq 6\farcs0$) and include no rich
cluster lenses.

Keeton (1998), followed by Porciani \& Madau (2000) and Kochanek \& White (2001),
demonstrated that the origin of the problem lay in neglecting the baryonic physics
which makes the density structure of the lenses depend strongly on the mass scale.
Any model based on the halo mass function which assumes that the density distributions
of the halos vary smoothly and continuously with mass leads to predictions for
the image separation distribution which catastrophically fail to match the data.
Keeton (1998) demonstrated it for singular isothermal sphere (SIS) and NFW
models, while Kochanek \& White (2001) demonstrated it for the adiabatically
compressed models described in \S2 and illustrated in Fig. 1.  
The key, which can be understood self-consistently based on the adiabatic 
compression models described in \S2 (Kochanek \& White 2001), is that the
lensing efficiency of a halo increases dramatically as the halo is compressed by 
the cooling baryons.  For example, Fig. 2 shows the change in the
deflection profile (bend angle) for a $10^{12}M_\odot$ halo at $z_l=0.5$ with a 
source at $z_s=2$ between the initial NFW model and the final adiabatically
compressed model with mass fraction $m_d+m_b=0.05$ in cold baryons.  Where
the initial NFW halo is sub-critical and unable to generate multiple images,
the compressed halo is super-critical.  The final cross section depends strongly
on the details of the baryon distribution, as the models with bulge-to-disk mass
ratio of $m_b/m_d=20\%$ are significantly better lenses than those without
a bulge component.  

Keeton (1998) and Porciani \& Madau (2000)  phenomenologically solved the problem of 
determining the global separation distribution by breaking the mass
function into a high mass (cluster) distribution modeled by standard NFW profiles
and a low mass (galaxy) distribution modeled using the local properties of galaxies
rather than the theoretical mass function.  In order to fit the observed separation
distribution, the break had to be located at a cooling mass scale near 
$M_c = 10^{13} M_\odot$.  Given the difficulty in performing ab initio calculations
for the final density structure of galaxies (as illustrated by the strong 
dependence of the bend angles on the bulge mass fraction in Fig. 2), such
phenomenological methods are likely to be more quantitatively useful than
current attempts at ab initio calculations.

While recognizing this limitation on the quantitative use of current theoretical
models for the final density structure of galaxies, Kochanek \& White (2001) used the
theoretical models to illustrate and isolate the physics needed to produce the
observed separation distribution from the halo mass function.  They first
assumed that for halos less massive than a cooling mass scale $M_c$, baryonic
mass fraction $m_d+m_b=0.05$ of the halos cools.  Fig. 3 shows the predicted
separation distributions as a function $M_c$.  Recall, from Fig. 1, that if
either all (high $M_c$) or no (low $M_c$) halos cool, then we cannot match 
the observations.  Only for a cooling scale $M_c \sim 10^{13} M_\odot$ can
we produce a sharp break in the separation distribution on the $\Delta\theta=3\farcs0$ 
scale of the observations. The exact scale depends on the density structure
of the cooled halos, with $M_c$ decreasing from $10^{13}M_\odot$ to 
$5\times 10^{12} M_\odot$ when we add a $m_b/m_d=0.1$ bulge.  Interestingly,
cosmological hydrodynamic simulations also find that the cooled baryon
fraction reaches 50\% on mass scales near $10^{13}M_\odot$ (e.g. Pearce et al. 1999). 
The required cooling mass scale also depends on the cold baryon fraction, with
$M_c$ decreasing as $\log_{10} M_c/M_\odot\simeq 13.6-(m_d+m_b)/0.15$ when the
cold baryon fraction $m_d+m_b$ increases.  The Kolmogorov-Smirnov (K--S) test
probability of fitting the observed separation distribution is shown in 
Fig. 4 for a range of cold baryon fractions.

These two parameters,
the cooling mass scale and the cold baryon fraction, are not independent.
The cooling mass scale $M_c$ at any epoch is the mass scale where the cooling
time is roughly equal to the average age of the halos.  Fig. 4 also shows
that for the current epoch, the two time scales are equal near 
$M_c=10^{13}M_\odot$ for $m_d+m_b=0.05$.  A lower cold baryon fraction
requires a mass scale $M_c$ for which there is insufficient time for the
baryons to cool.  A higher cold baryon fraction requires mass scales where
there is too much time for the baryons to cool.
Since (to first order!) cooling physics combined with the baryonic mass fraction determines
$M_c$, the fundamental physical parameter leading to the observed separation
distribution is the cosmological baryon density $\Omega_b$.  If all baryons
cooled and remained cold, then there would be no ambiguities to this 
statement.  However, star formation and feedback can reheat large fractions
of the baryons even if they cool initially (see SA), so that the cosmological
density in cold baryons, $\Omega_{b,{\rm cool}}$, can be significantly less than
the total density in baryons $\Omega_b$.  In our simple models we assume
that halos start with (eventually cold) baryon mass fraction
$(m_d+m_b)_0=\Omega_{b,{\rm cool}}/\Omega_m \leq \Omega_b/\Omega_m=0.13$ and that
the fraction which has cooled and compressed a halo of mass $M$ at redshift 
$z$ is $m_d+m_b = (m_d+m_b)_0 f_{cool}(M,z)$ where the simple model of
the cooling function outlined in \S2 determines $f_{cool}(M,z)$.  Fortunately,
the estimates of $\Omega_{b,{\rm cool}}$ depend very weakly on changes in 
$f_{cool}$, the lens sample used, or the addition of a small bulge component.

The K--S test probability of fitting the observed separation distribution
as a function of the cold baryon density $\Omega_{b,{\rm cool}}$ is shown
in Fig. 5.  With little sensitivity to the details, models with
 $0.015 \ltorder \Omega_{b,{\rm cool}} \ltorder 0.025$ agree with
the data.  While the preferred range is less than the total baryon density
$\Omega_b=0.04$ in the input cosmology, it significantly exceeds
the estimates of $0.0045 \ltorder \Omega_{b,{\rm cool}} \ltorder 0.0068$
for the cold baryon fraction (stars, cold gas and stellar remnants) in
local galaxies by Fukugita, Hogan \& Peebles~(\cite{Fuk98}).  This is
a weighted average over all galaxies rather than simply that of massive
galaxies, but the conflict is probably present even for massive galaxies
like the Milky Way.  Models of the Galaxy by Dehnen \& Binney~(\cite{Dehnen98}) 
find a baryonic mass fraction $m_d+m_b \simeq 0.08$ corresponding to
$\Omega_{b,{\rm cool}} \simeq 0.024$ at 100~kpc.  As we increase the
radius we include additional dark matter but no new baryons, so the
baryon fraction drops to $0.008 \ltorder \Omega_{b,{\rm cool}} \ltorder 0.012$
for a halo with an outer extent of 200--300~kpc.  While the difference
is smaller, it suggests the existence of a {\it dynamical baryon discrepancy},
in which the cold baryon fraction required to explain the observed dynamical 
properties of galaxies exceeds standard accountings for the cold baryons in 
galaxies. 

\section{The Local Velocity Function} 

The distribution of gravitational lens image separations is the cleanest dynamical
probe of the halo mass function because the image separation is directly related 
to the underlying halo density distribution and because lensing selects halos 
without any dependence on the luminosity of the baryons.  Unfortunately, while the 
number of known lenses is growing relatively rapidly, the overall size of the 
sample is limited. Our
alternative dynamical probe of the mass function is the local velocity function
of galaxies and clusters.  The local velocity function has the opposite problems
from the lenses, with far lower statistical uncertainties but far higher systematic
uncertainties.  First, selection methods for galaxies and clusters
depend on the luminosity and surface brightness of the halo.  Second, the 
selection methods for galaxies and groups/clusters are inhomogeneous, leading
to significant systematic difficulties when assembling a global velocity
function incorporating both. Third, local dynamical probes of galaxies and
clusters have many more ambiguities than gravitational lens image separations
because they are only indirectly related to the mass distribution.  This problem
is worst for galaxies where there are significant systematic uncertainties in
the relationship between stellar kinematics and the underlying mass distribution.

Most estimates of the local velocity function of galaxies have been made because
it is an essential element in estimates of gravitational lens cross sections,
separation distributions, and related statistics (e.g. Turner et al. 1984,
Fukugita \& Turner 1991, Kochanek 1993, 
1996; Maoz \& Rix 1993; Falco et al. 1998; Helbig et al. 1999).  Cole \& Kaiser
(1989) pointed out that the velocity function can also be used to test models of
galaxy formation, the power spectrum and cosmology directly.  This motivated 
estimates of the local velocity function by Shimasaku (1993) and Gonzalez et al.
(2000) and further theoretical investigations by Sigad et al. (2000), Newman
\& Davis (2000), and Bullock et al. (2001).  All derivations of the galaxy velocity function
use the Faber-Jackson (1976) and Tully-Fisher (1977) relations between luminosity
and velocity to perform a variable transformation from a locally measured 
luminosity function into the velocity function. Far more effort has focused
on determining the mass or velocity (temperature) function of groups and clusters
(see the reviews by Rosati \& Donahue in these proceedings) and given the greater 
familiarity of these results we will
not review them in any detail.  It is difficult to merge the two, to produce
a global view of the mass function, because of the problems in obtaining 
complete, well-understood samples of groups and then determining their masses
or velocities (e.g. Mahdavi et al. 2000).

Existing observational estimates of the velocity function of galaxies suffer from (at
 least!) five 
systematic problems.  First, deriving a velocity function requires the luminosity
function of galaxies by type since the kinematics of pressure-supported early-type
and rotation-supported late-type galaxies are very different.  Unfortunately, there
are large systematic differences between morphologically-typed and spectrally-typed
luminosity functions.  However, in Kochanek et al. (2001b) we find clear evidence that
current spectrally-typed luminosity functions have internal inconsistencies which
make the morphologically-typed surveys the better option at present.  Second, the
luminosity functions and the kinematic relations used to construct the velocity 
function are derived on different magnitude scales, leading to systematic shifts in
the velocity scale.  Third, by separating the derivations of the luminosity function
and the kinematic relations, covariances which are important to the uncertainties in
the velocity function are lost.  Fourth, the classification method used for the
luminosity function (morphological or spectral) is not the same as the kinematic
classification (pressure or rotation supported) used to define the kinematic relations.
The galaxies found in kinematic samples do not even represent fair samplings of the
classifications used to construct the luminosity function as they are dominated by
E and Sb/Sc galaxies with few S0/Sa galaxies.  Fifth, the kinematic velocities have
specific observational definitions that are not easily translated into theoretically
calculated quantities.  For example, the central stellar velocity dispersion of an
early-type galaxy is non-trivially related to the peak circular velocity of a model
halo.

\begin{figure}
\centerline{\psfig{figure=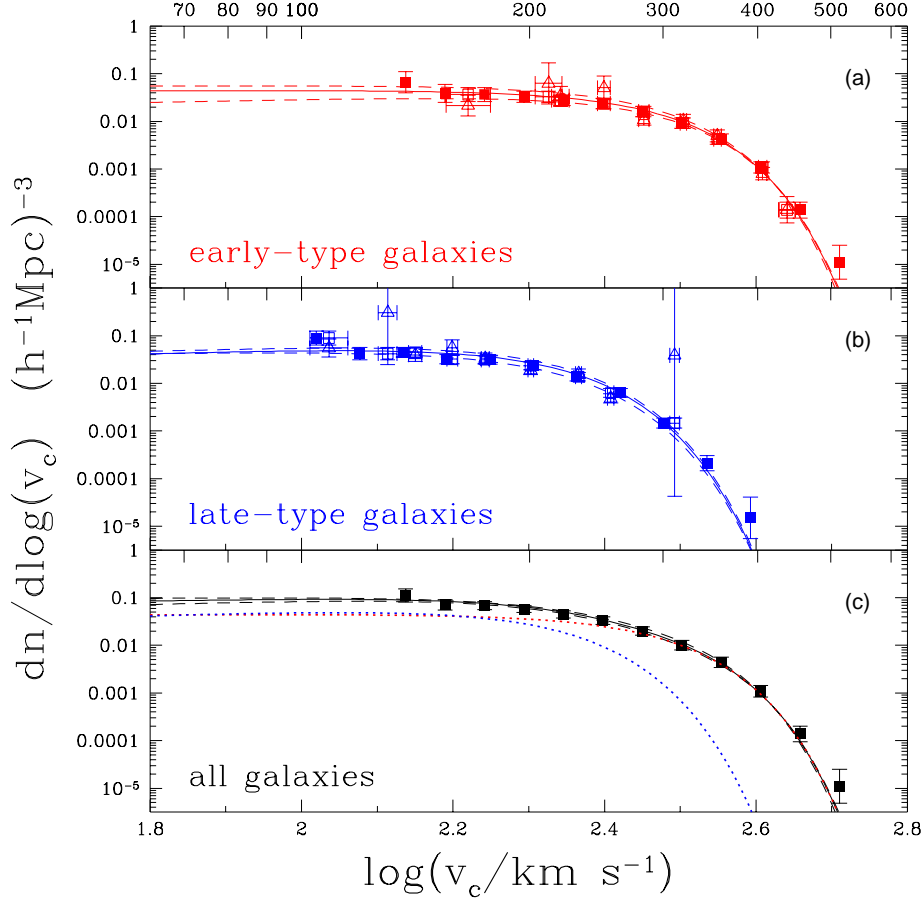,width=5.0in}}
\caption{ (a)~The early-type galaxy velocity function.  The curves show our parametric
estimates of the velocity function.  The solid curve is our standard estimate with the
galaxy type boundary at $T=-0.5$, while the upper (lower) curves show that the number of
low velocity early-type galaxies is strongly affected by shifts in the type boundary
to $T=+0.5$ ($-1.5$).  The points show the three different non-parametric estimates of
the velocity function derived from the binned, non-parametric estimate of the luminosity
function.  The solid squares use the analytic Faber-Jackson relation to transform both
the magnitude and the density, the open squares use it only to transform the density,
and the open triangles do not use it at all.  (b)~The late-type galaxy velocity function.
The curves and symbols are the same as in (a).  (c)~The total galaxy velocity function.
The curves are the same as in (a) and (b), but now calculated for all galaxies
simultaneously.  The individual early- and late-type galaxy
contributions to the velocity function are plotted as dotted lines.
  }
\end{figure}

\begin{figure}
\centerline{\psfig{figure=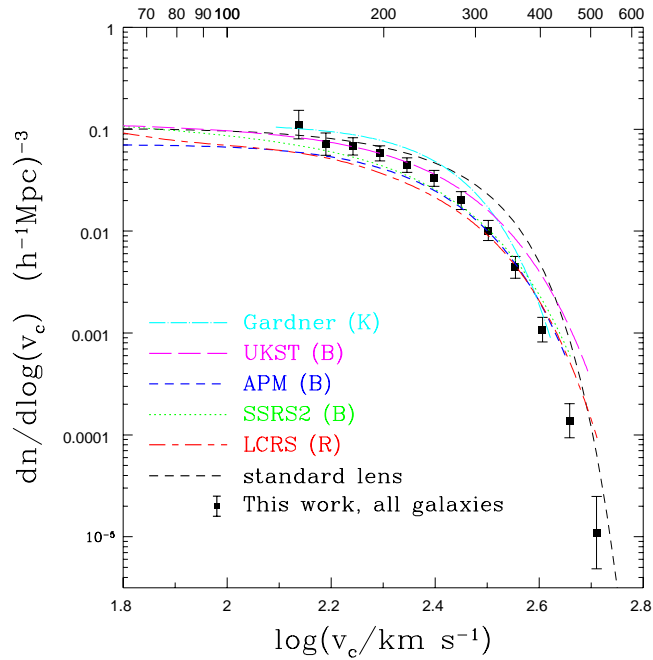,width=3.5in}}
\caption{ The total galaxy velocity function compared to previous estimates from
  Gonzalez et al. (2000) and Falco et al. (1998).  The standard lens model can
  be offset in velocity because of differences in the observational method
  (see \S5).  }
\end{figure}

\begin{figure}
\centerline{\psfig{figure=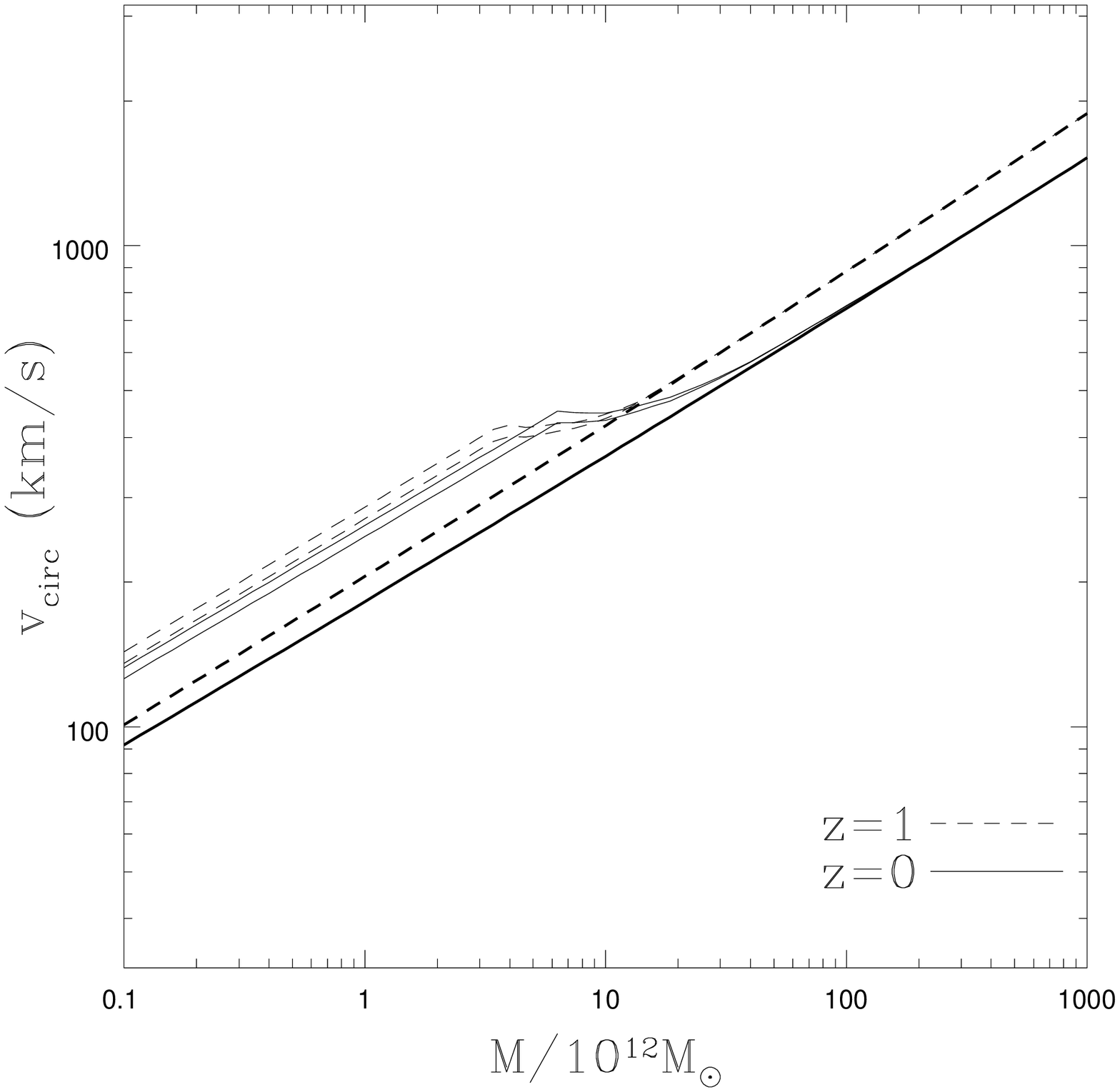,width=3.3in}}
\caption{The global relation between mass and circular velocity at redshifts
zero (solid) and unity (dashed).  The heavy curves show the peak circular
velocity of the NFW model.  The light curves show the peak circular velocity
including the baryonic cooling and adiabatic compression from the
$\Omega_{b,{\rm cool}}=0.018$ model.  The upper light curve is
the model with no bulge component ($m_b/m_d=0$) and the lower light
curve is the model with a bulge ($m_b/m_d=0.1$).  The bulge slightly
reduces the peak rotation velocity (because it increases the angular
momentum per unit mass of the disk) while making the rotation curve
flatter. }
%\end{figure}
%\begin{figure}
\centerline{\psfig{figure=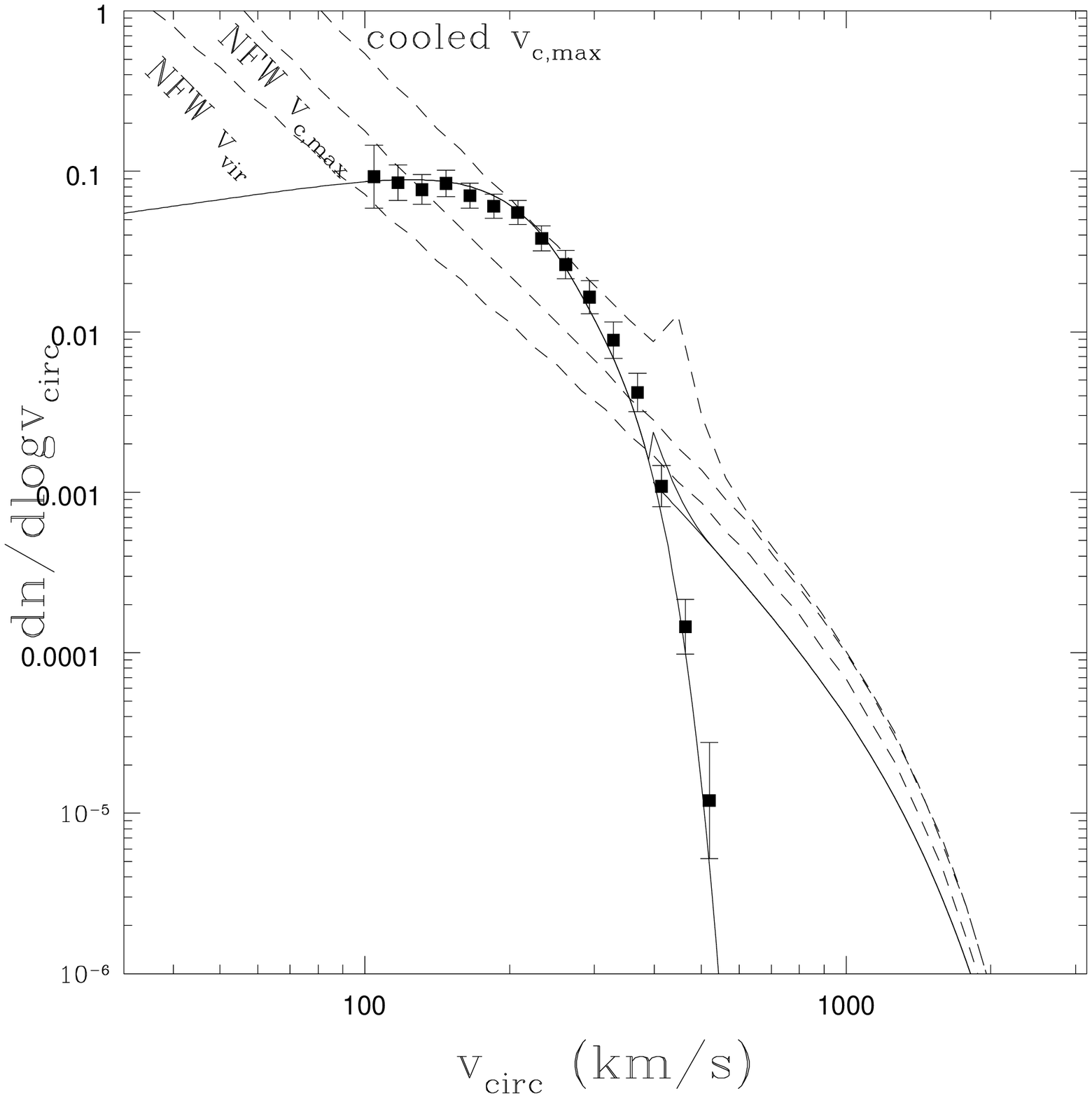,width=3.3in}}
\caption{The velocity function
$dn/d\log v_c = (dn/d\log M)|d\log M/d\log v_c|$.
The solid curves show the local velocity function of galaxies (low $v_{circ}$)
and clusters (high $v_c$) and their sum.
The points are the non-parametric velocity function of galaxies.  From bottom
to top, the dashed curves show the velocity functions derived using $dn/dM$
and the NFW virial velocity (labeled NFW $v_{\rm vir}$), the peak circular
velocity of the NFW rotation curve (labeled NFW $v_{c,{\rm max}}$) and the
peak circular velocity of the adiabatically compressed model (labeled cooled
$v_{c,{\rm max}}$).  We used the $\Omega_{b,{\rm cool}}=0.018$ model with no
bulge.}
\end{figure}

In Pahre et al. (2001) we took three steps to reducing these systematic uncertainties.
First, we started from a large, local infrared luminosity function (Kochanek et al. 
2001a) derived from the 2MASS (Skrutskie et al. 1997) survey.  The galaxies were
morphologically classified and the classifications are internally consistent and
consistent with other local, morphologically classified samples (Kochanek et al. 2001b).
We then matched galaxies in the 2MASS sample to several modern kinematic surveys to
construct Faber-Jackson and Tully-Fisher relations in the same magnitude system
as was used to derive the luminosity function.  We then combined the luminosity
function and the kinematic relations including the full variable covariances of
the functions to determine the velocity function.  We also explored non-parametric
models for the velocity function where we minimized the use of functional forms
for the luminosity function or the kinematic relations in favor of the raw,
binned data.  Fig. 6 shows the resulting velocity functions, and Fig. 7 compares
our velocity function to other derivations (from Gonzalez et al. 2000) and to
the standard result from gravitational lens statistics (Falco et al. 1998). While
our result has reduced systematic errors compared to earlier derivations, we have
not eliminated problems created by the difference between morphological and 
kinematic types or the problems in relating observed and theoretical velocities.
 
Our theoretical model, tuned to reproduce the distribution of gravitational lens
separations in \S3, also estimates the peak circular velocity as a function of
mass, as shown in Fig. 8.  Below the cooling mass scale, the baryons compress
the halos and shift the mass-velocity relation $v_c(M)$ upwards.  The amount of
the shift depends on both the assumed baryon distribution (as illustrated by
comparing models with bulge-to-disk mass ratios of $m_b/m_d=0$ and $0.1$) 
and redshift.  With added physics (such as the reheating mechanisms in the
SA models), the simple power-law structure below the cooling mass scale 
changes.  We can use the mass-velocity relation to derive the velocity function
from the mass function as a simple variable transformation (eqn. 6.2 with
$p_g(M)=1$) with the results shown in Fig. 9.  We used three $v_c(M)$ relations: the virial
velocity of the NFW halos, the peak circular velocity of the NFW halos, and the
$v_c(M)$ relation produced by the model which best fit the gravitational lenses.
We also extended the observed velocity function by adding a crude estimate for
the contribution of groups and clusters to the local velocity function. We 
estimated the cluster contribution from the Blanchard et al. (2000) X-ray 
temperature function combined with the Wu et al. (1999) relation between the 
temperature and the velocity dispersion (a simple thermodynamic conversion is
nearly identical) truncated to $T > 0.5$~keV.  This produces the small kink in
the estimate near $v_c=400\kms$.  On cluster scales the models and the data
agree relatively well, although a modestly lower normalization for the mass
function would fit the data better.  On galaxy scales, the models without
the compression by the baryons grossly under predict the density of halos 
with observed velocities $200 \kms \ltorder v_c \ltorder 400\kms$ while the
models with adiabatic compression match the density relatively well.  

Our model clearly has problems at low velocity and at the juncture between
galaxies and groups/clusters.  These are not apparent in the models for the
gravitational lenses because the lens cross sections combined with the 
angular selection functions strongly suppress the contributions from halos
with $v_c \ltorder 150\kms$, and because the separation distribution 
smoothes the velocity function by an average
over lens redshift.  The peak in the model distribution near $400\kms$ is
an artifact of the flat slope of the $v_c(M)$ curves (see Fig. 8).  The 
$v_c(M)$ relation has to be multi-valued in this region with
a diminishing probability of forming a compressed galaxy halo and a rising
probability of forming an uncompressed group halo (we will explore this
further in \S6).  Gonzalez et al. (2000), who made a similar comparison
to the velocity function of galaxies based on the Somerville \& Primack (1999) 
semi-analytic models, had very similar problems. Even though they used
a model where feedback from star formation varied the cold baryon fraction
to maximize the compression at $v_c \sim 250\kms$ and to reduce it
for higher and lower velocity halos, their model velocity function more
closely resembles our theoretical curves than the observations.

\section{Deriving the Velocity Function From the Lens Separations} 

\def\dt{\Delta\theta}
\def\dth{\Delta\hat{\theta}}

As a final check of the consistency of the model, the distribution of lens separations
and the velocity function, we can estimate the velocity function directly from the
distribution of lens separations so that the comparison does not depend on our
theoretical model from \S2.  We will compare only the shapes of the velocity
function and not the absolute normalization (number per comoving Mpc) since
the normalization introduces the uncertainties in the absolute numbers
of gravitational lenses found by a survey.  We assume that lenses can be 
modeled as singular isothermal spheres (SIS), which is broadly consistent
with both lensing and dynamical data on the relevant scales (see Cohn et al. 2001), so that 
the observed image
separation is a simple function of the circular velocity $v_c$ and the 
lens-source/observer source distance ratio, 
$\Delta\theta = \Delta\theta_0 (v_c/v_0)^2 D_{LS}/D_{OS}$
where $\Delta\theta_0=4\pi (v_0/c)^2$ sets an arbitrary velocity scale 
for the calculation.  In any flat cosmology, the normalized image 
separation distribution is
\begin{equation}
 { 1 \over \tau } { d \tau \over d\dt } =
 { \dt^2 S(\dt) \over \dt_0 }   
 \left[ \int_0^1 x^2 dx \int_0^\infty dv \dt^2 S(\dt) { dn \over dv} \right]^{-1}
   \int_{v_{min}}^\infty dv
    { v_0^2 \over v^2}{dn\over dv}\left( 1 - { v_0^2 \over v^2 }{\dt \over \dt_0} \right)^2 
\end{equation}
where $v_{min}=v_0(\dt/\dt_0)^{1/2}$,
$0 \leq S(\dt) \leq 1$ is the survey selection function for finding a lens
of separation $\dt$, $x=D_{OL}/D_{LS}$, and $dn/dv$ is the velocity function.  
We can non-parametrically determine the velocity function using a variant of
the step-wise maximum likelihood (SWML) luminosity function estimation method
of Efstathiou et al. (1988).  We approximate $dn/dv$ by a series of bins with
density $n_i$ and then maximize the likelihood of finding the observed 
lens separations while holding the number of lenses in the bin centered
on $v_c=300$~km/s fixed to unity.  The latter constraint corresponds to
ignoring the absolute comoving density of the lenses and considering only
the shape of the velocity function. Eqn. (5.1) holds for
any flat cosmological model.  

The results for three lens samples are shown in Fig. 10. Sample A consists
of the 20 lenses found in surveys based 8~GHz VLA A-array maps of flat-spectrum
radio sources by the CLASS and PMN (Winn et al. 2000) surveys.  Sample B,
with 27 lenses, adds the remaining radio-selected lenses.  Sample C, with
46 lenses, adds the optically-selected quasar lenses.  The model for the
angular selection function is adjusted for the angular sensitivity of
the survey which found each lens.   As we move from Sample A to C we trade
increasing systematic uncertainties for decreasing statistical uncertainties,
although we find that the derived velocity functions are mutually consistent
given the statistical uncertainties.  Only the lowest velocity bin,
centered at $v_c=100$~km/s and corresponding to an average image separation
of only $\langle\Delta\theta\rangle=0\farcs15$, has a significant sensitivity 
to plausible errors in the models for the angular selection function.  Be
warned that the error bars in Fig. 10 are even more highly correlated than
similar figures derived for luminosity functions of galaxies!

\begin{figure}
\centerline{\psfig{figure=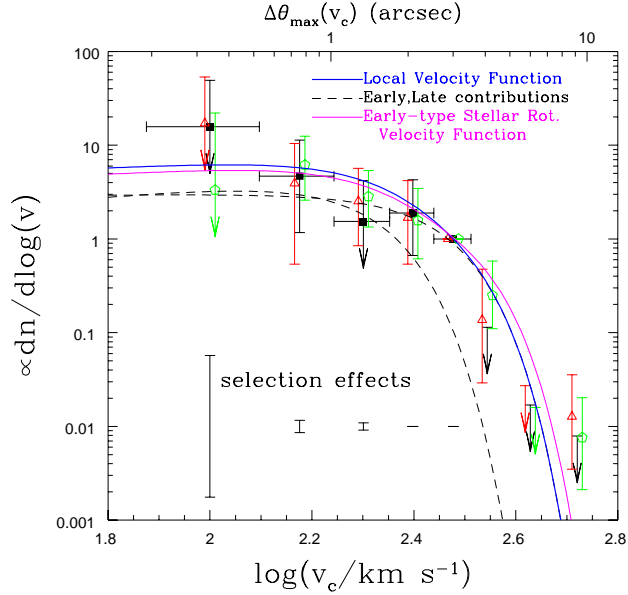,width=3.5in}}
\caption{ 
The velocity function estimated from gravitational lenses.  The solid points are derived 
using only the flat-spectrum radio surveys, the triangles include the MIT-Greenbank survey,
and the pentagons include quasar lenses.  The triangles and pentagons are slightly offset 
in velocity to make them more visible.  The horizontal error bars show the width of the 
velocity bins.  The distributions are normalized by the logarithmic density at 
$v_c=300\kms$.  The solid (dashed) line shows our locally estimated total (early-type and 
late-type) velocity function.  The error ranges labeled ``selection effects'' show the 
effects of plausible uncertainties in the angular selection function $S(\Delta\theta)$.  
The scale at the top of the figure shows the maximum image separation produced by a lens 
with circular velocity $v_c$.  The mean separation is one-half the maximum.  }
\end{figure}

If we normalize our estimate of the velocity function of galaxies at the
same velocity scale and superpose it on that of the lenses, the two
distributions are remarkably similar (see Fig. 10)  They have the same
flat low-velocity slope as a function of $\log(v_c)$ and an exponential
cutoff on the same velocity scale.  Only in the highest velocity bin,
whose density is driven by the need to produce the widest separation
Q~0957+561 lens with $\Delta\theta=6\farcs1$, does the velocity function
of the lenses show a clear deviation from that of galaxies even though
the lenses represent the global velocity function rather than that of 
the galaxies alone.  The small amplitude of the deviation is another 
illustration of the enormous impact of the baryons on the dynamical
structure of halos which we discussed in \S3. 
At low velocities the two distributions have the same flat slope, rather
than the steeply rising slope of our estimates from the mass function. 
Note, however, that neither observational sample has significant data
on velocity scales $v_c \ltorder 100\kms$ where the deviations from 
the predictions begin to diverge rapidly.

\section{A Non-Parametric Description of the Velocity Function}

Our theoretical model is a gross oversimplification.  It agrees with our two
observational probes on the mass scales corresponding to massive galaxies and
clusters, but has problems on the mass scales of groups and fails badly for
low mass galaxies.  These problems can be partially rectified by more 
sophisticated models (e.g. SA) which allow for more complicated variations in 
the cold baryon fraction with halo mass.  In this section we outline a general, 
non-parametric approach to understanding the relationship between the mass function 
of halos and the velocity function of galaxies which we can use to characterize
the problem without the complications of a full semi-analytic model.  

Our starting point is that in all models, no matter their complexity, the halo
mass function and the galaxy velocity function are related by the probability
$P(v_c|M)$ that a halo of mass $M$ forms a detectable galaxy with
circular velocity $v_c$,
\begin{equation}
  { dn \over dv_c } = \int_0^\infty { dn \over dM } P(v_c|M) dM.
\end{equation}
The conditional probability, which need not integrate to unity, includes the effects
of all parameters governing the formation of galaxies such as the spin parameter
$\lambda$, the collapse redshift, the halo merger history and its environment.
If $P(v_c|M)$ is dominated by a sufficiently narrow ridge, so that it is reasonable
to associate a characteristic velocity with halo mass, then we can
approximate the integral (6.1) by
\begin{equation}
  { dn \over dv_c } (v_c(M)) =
     p_g(M) \left| { d v_c(M) \over d M } \right|^{-1} { dn \over dM}
\end{equation}
where the unknown two-dimensional function is replaced by two one-dimensional
functions with simple physical meanings.  The first, $p_g(M)$, is the probability
that a halo of mass $M$ forms a galaxy included in the velocity function, and
the second, $v_c(M)$, is the circular velocity of the resulting galaxy. As
a mathematical derivation we must assume that the fractional spread in velocity
at fixed mass, $\sigma_v(M)/v_c(M)$, is small.  Since the equivalent fraction
at fixed luminosity is indeed small, and changing from mass to luminosity
presumably raises rather than lowers the dispersion in galaxy properties, it
seems likely that the expansion is justified.  This model differs from that
used in \S3 and \S4 where $p_g(M)\equiv 1$.  Its main drawback is that it
neglects ``halos-in-halos'' or the halo multiplicity function (e.g. Peacock
\& Smith 2000, Soccimarro et al. 2001), although this should be a modest
perturbation in the accounting for galactic halos ($\sim10\%$ in the
numerical simulations of White et al. (2001).
These two functions are implicitly included in semi-analytic models.  For example,
Gonzalez et al. (2000) fit the cold baryon fraction in the Somerville \&
Primack (1999) models by $m_d(x)=0.1(x-0.25)/(1+x^2)$ for
$x=v_{mod,0}(M)/200\kms > 0.25$ (see eqn. 2.1) to estimate $v_c(M)$ for
the model.  The velocity function
predicted by this model cannot, however, reproduce the local velocity
function of galaxies, as discussed in \S4 (see Fig. 9).

\begin{figure}
\centerline{\psfig{figure=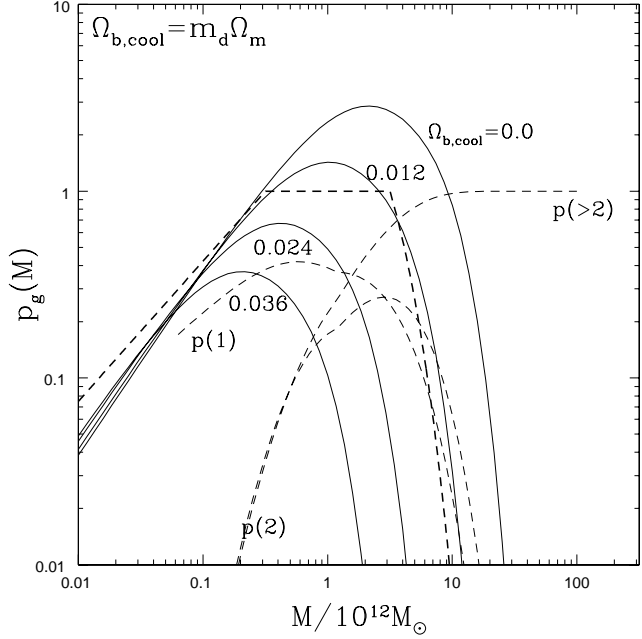,width=3.5in}}
\caption{The galaxy formation probability, $p_g(M)$, required for models with 
  constant cold baryon fractions.  The curves are
  labeled by their cold baryon density, $\Omega_{b,\rm{cool}}= m_d \Omega_m$.
  The upper curve is for halos with no cooled baryons ($\Omega_{b,\rm{cool}}=0$)
  and the lowest curve is for halos with $\Omega_{b,\rm{cool}}\simeq\Omega_b$.
  The heavy dashed curve is the standard model from our attempt to non-parametrically
  adjust $v_c(M)$ in order to maximize the mass range over which $p_g(M)=1$.
  The light dashed curves show the probability of forming one ($p(1)$), 
  two ($p(2)$) or at least two ($p(>2)$) galaxies as a function of 
  halo mass in one of the halo multiplicity models of Scoccimarro et al. (2001). 
  The Scoccimarro et al. (2001) models are low because they underestimate the 
  comoving density of galaxies (see text).
  }
\end{figure}

\begin{figure}
\centerline{\psfig{figure=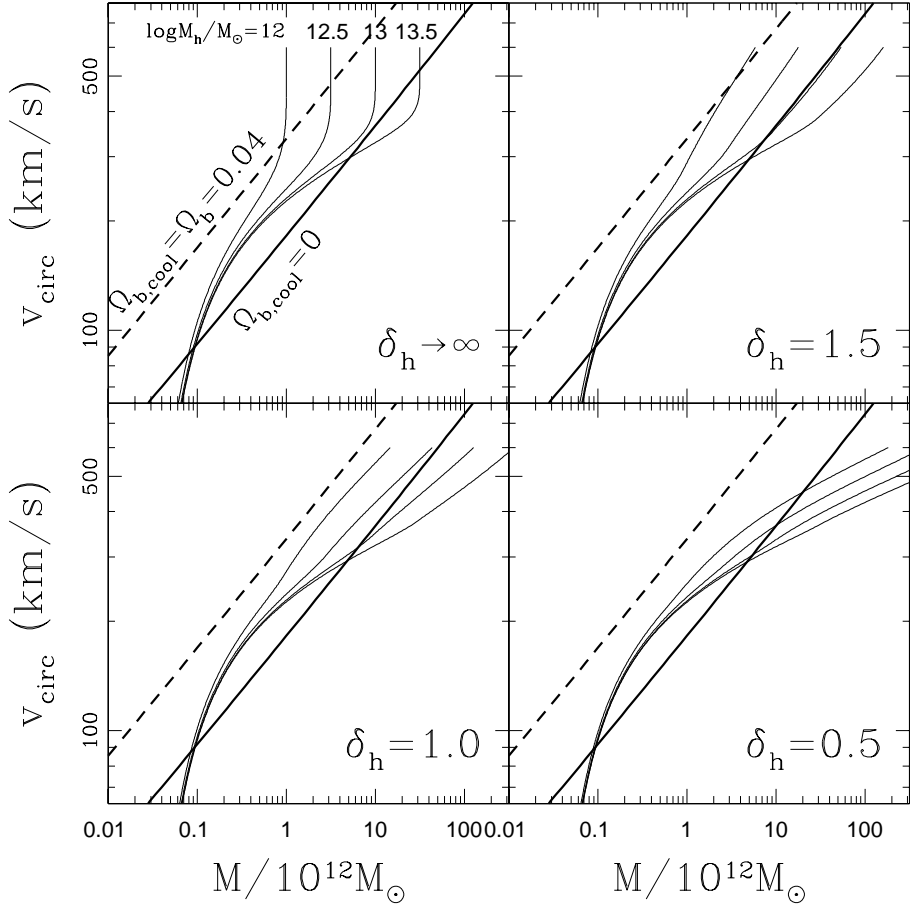,width=5.0in}}
\caption{ The effect of the high mass cutoff in the galaxy formation probability on
  the mass-velocity relation.  The four light curves in each panel show the mass-velocity
  relations $v_c(M)$ needed to produce the observed velocity function given high
  mass cutoffs in the formation probability of $\log M_h/M_\odot=12.0$, $12.5$,
  $13.0$ and $13.5$ (from left to right).  The different panels show the effect
  of changing the cutoff exponent, with values of $\delta_h\rightarrow\infty$
  (top left), $1.5$ (top right), $1.0$ (lower left) and $0.5$ (lower right)
  covering the range from an infinitely sharp cutoff to a fairly shallow cutoff.
  The heavy curves show the permitted range for $v_c(M)$ with a lower boundary
  set by the circular velocity of uncompressed halos (the heavy solid line)
  and the upper boundary set by the circular velocity
  of halos in which all the available baryons have cooled (the heavy dashed line)
  The upper bound, which assumes $\lambda=\bar{\lambda}=0.05$ and $j_d=m_d$, is
  significantly more model dependent than the lower bound.
   }
\end{figure}

\begin{figure}
\centerline{\psfig{figure=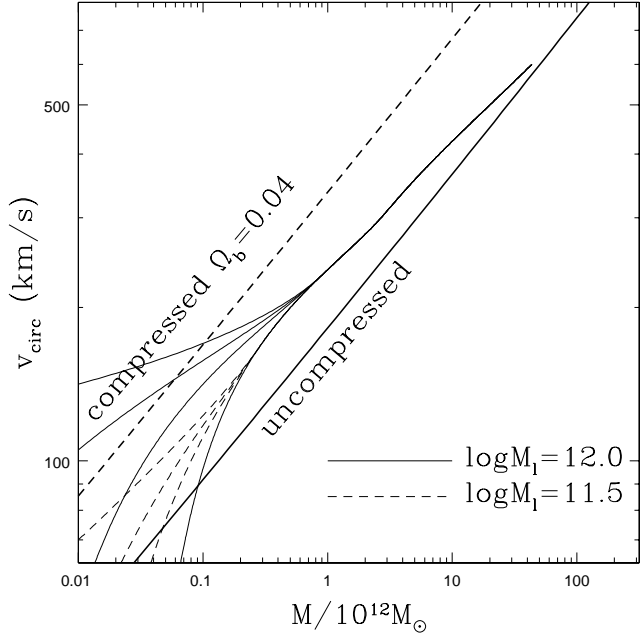,width=3.5in}}
\caption{ The effect of the low mass cutoff in the galaxy formation probability on $v_c(M)$.
  The light solid and dashed lines show the $v_c(M)$ relations needed to produce the
  observed velocity function given low-mass cutoffs at $\log M_l/M_\odot=11.5$
  (dashed) and $12.0$ (solid) for power law exponents of $\delta_l=1$ (top), $3/4$,
  $1/2$, and $0$ (bottom).  The case $\delta_l=0$ corresponds to having no low-mass
  cutoff and the two $M_l$ cases overlap. The upper cutoff is fixed to $M_h=10^{12.5}M_\odot$
  with $\delta_h=1.0$.  The heavy solid and dashed lines are the lower and upper
  bounds corresponding to the uncompressed and the maximally compressed limits of
  the model.
  }
\end{figure}

\begin{figure}
\centerline{\psfig{figure=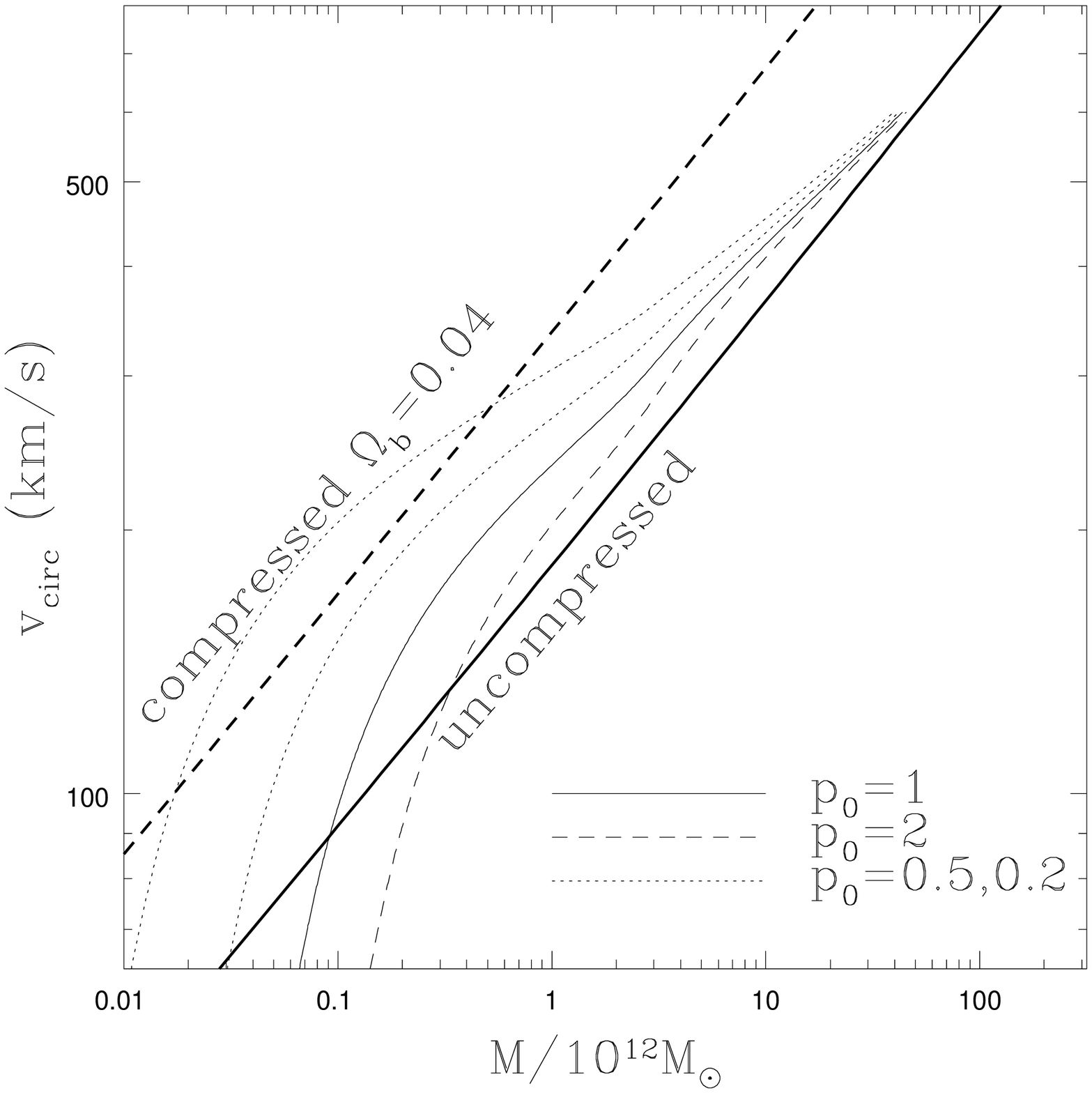,width=3.5in}}
\caption{The effect incompleteness or normalization errors on $v_c(M)$.
  The light curves $v_c(M)$ for $p_0=2$ (dashed), $p_0=1$ (solid),
  $p_0=1/2$ (dotted) and $1/5$ (dotted) using the standard high mass cutoff.
  The heavy solid and dashed lines are the lower and upper
  bounds corresponding to the uncompressed and the maximally compressed limits of
  the model. }
\end{figure}

Our decomposition of the problem into the formation probability $p_g(M)$ 
and the mass-velocity relation $v_c(M)$ allows us to explore the problem
in a model independent fashion. Unfortunately, the solution is not unique
because we must determine both $p_g(M)$ and $v_c(M)$ from only one
function.  Fortunately, the two unknown functions should obey several
constraints.  First, if the normalization of the mass function is 
correct, $0 \leq p_g(M) \leq 1$.  We can have $p_g(M)>1$ only if we interpret
it as a normalization error in the mass function or if the halo 
multiplicity function on galactic mass scales differs significantly
from unity.  Second, cooling
baryons should only increase the halo circular velocity, 
$v_c(M) \geq v_{mod,0}(M)$.  Third, the compression must be bounded by
models in which all the available baryons have cooled, 
$v_c(M) \leq v_{mod}(M,m_d=\Omega_b/\Omega_m,\lambda)$.  Note that this
upper bound is more model dependent than the lower bound.

Suppose that the standard adiabatic compression models are correct and
that the dimensionless parameters (cold baryon fraction, spin parameter
$\cdots$) are the same for all galaxies.  These parameters fix $v_c(M)$,
allowing us to calculate the galaxy formation probability required to
produce the galaxy velocity function from the halo mass function.
Fig. 11 shows the implied $p_g(M)$ for various cold baryon fractions 
($m_d=0$, $0.04$, $0.08$ and $0.12$) and a fixed spin parameter 
($\lambda=\bar{\lambda}$).  At low mass, $p_g(M) \sim M$ is needed
to match the steep slope of the mass function to the shallow slope
of the velocity function, and at high mass it has the exponential cutoff
cutoff of the velocity function.  
The peak probability and the corresponding mass scale decrease systematically
as we raise the cold baryon fraction.  If the compression is too small
($m_d \ltorder 0.04$ or $\Omega_{b,{\rm cool}} \ltorder 0.01$), we must increase
the normalization of the halo mass function (i.e. the power spectrum)
in order to avoid a region with $p_g(M) > 1$.  When the compression of
the halos is large ($m_d \gtorder 0.08$ or $\Omega_{b,{\rm cool}} \gtorder0.02$),
either the velocity function is incomplete or we must lower the normalization
of the mass function.  

While these models clearly simplify the structure of the mass-velocity
relation $v_c(M)$ by using models with fixed dimensionless parameters,
they do not greatly exaggerate the dominant role of the formation 
probability in producing the observed shape of the velocity function.
The requirement that a physical mass-velocity relation be bounded
by the zero compression and maximal compression models prevents 
us from shaping $v_c(M)$ to produce the velocity function with the
formation probability held constant over much more than one decade
in mass.  We can illustrate this by using a parametric model for
$p_g(M)$ with a constant formation probability $p_g=p_0$ between
$M_l < M < M_h$, a power-law $p_g=p_0(M/M_l)^{\delta_l}$ below $M_l$,
and an exponential $p_g=p_0\exp(1-(M/M_h)^{\delta_h})$ above $M_h$.
For any set of parameters for $p_g(M)$ we can derive the $v_c(M)$
required to produce the velocity function.

\begin{figure}
\centerline{\psfig{figure=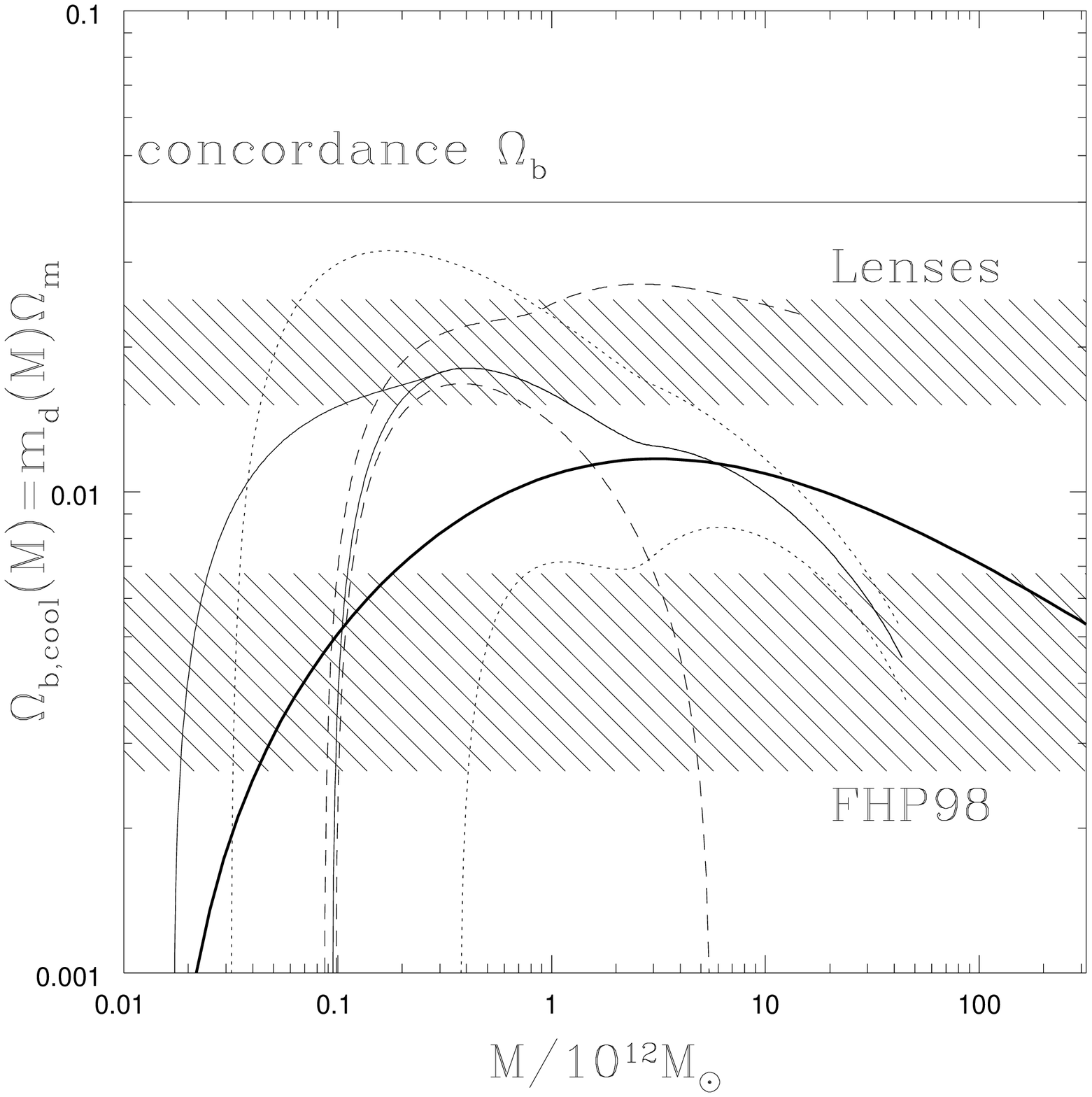,width=3.5in}}
\caption{Cold baryon densities ($\Omega_{b,{\rm cool}}=m_d \Omega_m$).  For a range of models of
  the galaxy formation probability, we use our adiabatic compression models with
  $\lambda=\bar{\lambda}=0.05$ and $j_d/m_d=1$ to convert $v_c(M)$ into an estimate
  for the cold baryon density.  The light solid curves are for a model with
  our standard high mass cutoff ($\log M_h/M_\odot=12.5$, $\delta_h=1$) with
  (upper curve at low mass) or without (lower curve at low mass) the standard
  low mass cutoff ($\log M_l/M_\odot=11.5$, $\delta_l=3/4$) in the formation
  probability.  The dashed lines show consequences of raising the upper cutoff
  to $\log M_h/M_\odot=13$ (bottom) or lowering it to $\log M_h/M_\odot=12$ (top).
  The dotted lines show the consequences of incompleteness ($p_0=0.5$, upper)
  and using too low a mass function normalization ($p_0=2$, lower) including
  only the standard high mass cutoff and no low mass cutoff.  The heavy solid 
  curve shows the cold baryon fraction estimate by Gonzalez et al. (2000)
  for the Somerville \& Primack (1999) semi-analytic models.
  The cold baryon density must be less than the total baryon density,
  $\Omega_{b,\rm{cool}} < \Omega_b$, and the total baryon density for the concordance
  cosmological model is marked with the upper horizontal line.  The hatched region
  labeled FHP98 shows the cold baryon density estimated by 
  Fukugita et al.~(1998) scaled to the concordance value of $H_0$.
  }
\end{figure}

Figs. 12--14 illustrate the effect of adjusting the structure of the formation 
probability on the mass-velocity relation $v_c(M)$.  We start with the high-mass 
cutoff ($M_h$ and $\delta_h$) since we are certain it exists in order to create
the distinction between galaxies and groups/clusters.  Physical models, where
$v_c(M)$ stays within its physical bounds, require a slightly blurred boundary 
with $\delta_h \simeq 1$ and $M_h \simeq 10^{12.5}M_\odot$ (which we adopt as 
our standard).  Steeper slopes $\delta_h$ allow modestly higher mass scales 
$M_h$ but the mass where $p_g(M)=1/2$ stays roughly in the range 
$10^{12.5}M_\odot$ to $10^{13}M_\odot$.  The group velocity function is set
by $1-p_g(M)$ with a velocity set by that of uncompressed halos, 
$v_c(M)=v_{mod,0}(M)$.  It is the lack of this multi-valued region in our
models from \S2 which leads to the ``kink'' in our model of the velocity
function near $v_c \sim 400\kms$ (see Fig. 9). 

Without a low-mass cutoff in the formation probability, $v_c(M)$ would have to rise
exponentially with mass in order to convert the steep slope of the mass function into
the shallower slope of the velocity function (see Fig. 9).  Since this rapidly 
requires velocities well below that of the halos, we are forced to introduce a
low-mass cutoff ($M_l$ and $\delta_l$) to the formation probability.  The effect
of changing various parameters is shown in Fig. 13, and we adopt 
$M_l = 10^{11.5}M_\odot$ and $\delta_l=3/4$ as our standard model.  All models
which allow $v_c(M) > v_{mod,0}(M)$ down to $30\kms$ have a formation probability
near 50\% at $10^{11}M_\odot$ and 10\% near $10^{10} M_\odot$.  We introduced
these models for $p_g(M)$ to see if we could produce a broad mass range in which
the formation probability was unity.  Fig. 11 superposes the standard model
we derived on the models for $p_g(M)$ found without any complicated structure
to $v_c(M)$.  The structure of the functions is nearly identical -- the best
we can manage is to keep the formation probability unity over one decade in
halo mass. From this we conclude that the  structure of the velocity function is  
dominated by the probability $p_g(M)$ of forming (or finding) a galaxy rather than 
variations in the relationship between halo mass and galaxy circular velocity
$v_c(M)$.  

Finally, in Fig. 14 we explore the effects of varying $p_0$, which models either 
errors in the normalization of the mass function or the completeness of the survey
underlying the velocity function.  We use the standard high mass cutoff and no
low mass cutoff.  For $p_0 <1$, which corresponds to making the velocity function
survey incomplete or lowering the normalization of the mass function, the velocity
corresponding to a given halo rises, and for $p_0 >1$, which corresponds to raising
the normalization of the mass function, the velocity falls. From this we can infer
that the velocity function galaxy sample cannot be massively incomplete and
the halo multiplicity function cannot differ greatly from unity.  

All these conclusions are very similar to the results from attempts to estimate the
halo occupation numbers needed to match the halo distributions found in simulations
to the observed distribution of galaxies (e.g. Peacock \& Smith 2000, Scoccimarro
et al. 2001).  For example, Fig. 11 also shows the probability of a halo forming
one, two or many galaxies in one of the models by Scoccimarro et al. (2001) 
based on semianalytic models and matching to large scale structure.  The two
distributions are very similar in structure given that the Scoccimarro et al. (2001) 
model has a different normalization.\footnote{When using the Scoccimarro et al. (2001) 
models to populate the halos from an N-body simulation of the cosmological model in \S2 
with galaxies, we found that they underestimated the galaxy density by factor of $\simeq 3$.} 
Nonetheless,
these more sophisticated models for the division of halos into galaxies would
greatly improve our more qualitative exploration of the problem. 
  
We finally return to the problem of the dynamical baryon discrepancy.  In Fig. 15
we convert some of the allowed mass-velocity relations into estimates of the 
cold baryon fraction based on the adiabatic compression model scalings 
(eqns. 2.1 and 2.2).  We also show the equivalent relation Gonzalez et al. (2000)
derived from the Somerville \& Primack (1999) semi-analytic models.  We
immediately see that the discrepancy between the estimated baryon content
of galaxies (from Fukugita et al. 1998) or the Galaxy (based on the models
of Dehnen \& Binney 1998) and the cold baryon 
fraction needed to compress the halos is present whether we use our simple
model of \S2 and the separation distribution of lenses, full semi-analytic
models, or totally non-parametric models.  

\section{The Future}

In this review we examined the relationship between the halo mass function and
dynamical probes of it using either the kinematics of galaxies or the separation
distribution of gravitational lenses.  Although many of the results we discuss
are implicit in full semi-analytic models, the approach of examining only the
dynamical properties has the advantage of eliminating the dependence of the 
comparison on luminosity.  Dynamical comparisons emphasize the critical 
role of the cooling baryons in transforming the dynamical structure of halos.
The baryonic compression produces a feature in both the distribution of
gravitational lens image separations and the local velocity function which
is a direct probe of the cold baryon fraction compressing galactic halos.
In all our comparisons we find a factor of 2--3 discrepancy between the mass
fraction in cold baryons required to explain the data and typical estimates
for the cold baryons in galaxies.  In a universe with $\Omega_b \simeq 0.04$
we need $0.01 \ltorder \Omega_{b,{\rm cool}} \ltorder 0.02$ to compress the halos
while typical accountings in galaxies find only 
$0.005 \ltorder \Omega_{b,{\rm cool}} \ltorder 0.01$ in known baryonic
components.  The difference could be explained by MACHOS (see Alcock, Richer and Sahu
in these proceedings), cold molecular gas or even warm gas in some circumstances.  
It is a part of our general problem that we lose track of most of the baryons at 
redshift zero (see Tripp in these proceedings).

There are three possible solutions to the dynamical baryon discrepancy. First,
it could be imaginary -- if you select your preferred ranges appropriately
you can essentially eliminate the discrepancy.  Second, it could be a problem 
in our models.  The adiabatic compression models are crude approximations for the
transformation of the dark matter halos by the baryons, and our models have
not properly accounted for the halo multiplicity function.  There is also a
major debate at present about the consistency of observed rotation curves
with the predictions of these standard models (see Burkert, Sancisi and
Sanders in these proceedings).  It is difficult to adequately address this 
possibility, since it is currently impossible to compute the final structure of 
a galaxy starting from the initial halo properties without approximations.
Third, the accounting for the baryons in galactic halos may be incorrect.
The Fukugita et al.~(\cite{Fuk98}) accounting for the baryons in galaxies
included only cold gas and normal stellar/stellar remnant populations,
neglecting hot gas ($10^6$~K), warm ionized gas ($10^4$--$10^5$~K),
and sub-luminous objects (e.g. MACHOS).  While hot gas cannot contribute to the 
adiabatic compression of the halo, the warm components are both difficult to 
detect and contribute to the compression.   

The most ambitious proposal for the future of the galaxy velocity function is to
use its evolution to determine the cosmological equation of state (Newman \& Davis
2000).  The challenge here is to understand, control, and then eliminate all the
sources of systematic error which can mimic or bias the evolutionary effect. 
Unfortunately, the evolution is subtle and achieving this goal will be difficult.
To put these difficulties in some context, many of the sources of systematic uncertainties 
are identical to those in the (currently unpopular) attempts to determine the 
cosmological model using gravitational lens statistics.
We can see some of the difficulties in our local comparisons between the halo
mass function and the velocity function of galaxies.  Different routes to
deriving the local velocity function (the choice of surveys, luminosity functions,
kinematic relations, type distributions $\cdots$), different models for the 
baryonic mass distribution in galaxies (the distribution of bulge-to-disk ratios,
cold baryon fractions, spin parameters $\cdots$) and different models for the
connection between halos and galaxies (formation probability, halo multiplicity
function $\cdots$)  all lead to differences that
are comparable to the effects of evolution.  Our non-parametric analysis shows
that at least two one-dimensional functions of the halo mass are needed, the
probability $p_g(M)$ that a halo of mass $M$ forms a galaxy included in the survey,
and the average velocity $v_c(M)$ of the resulting galaxy, and both of these
functions will themselves be evolving and survey-dependent.
However, studying the dynamical evolution of galaxies will yield so much
information on the evolution of galaxies, that the experiment will be of enormous
value even if systematic problems ultimately prevent it from being used for determining the
properties of the background cosmology.

\bigskip
\noindent Acknowledgments:  Much of this work was done in collaboration with M.
White (\S2 and \S3) or M. Pahre and E. Falco (\S4 and \S5) whose contributions
and comments were invaluable.  CSK is supported by the Smithsonian Institution 
and NASA grants NAG5-8831 and NAG5-9265.


\begin{thebibliography}{}
\bibitem[1996]{BauColFre}
  Baugh C.M., Cole S., Frenk C.S., 1996, MNRAS, 283, 1361
\bibitem[2000]{BCFBL}
  Benson A.J., et al., 2000, MNRAS, 311, 793
\bibitem[2000]{Blan00}
  Blanchard A., Sadat R., Bartlett J.G., Le Dour M., 2000,
  A\&A, 362, 809 
\bibitem[1986]{Blum86}
  Blumenthal G.R., Faber S.M., Flores R., Primack J.R., 1986, ApJ, 301, 27
\bibitem[2000]{Bro00}
  Browne I.W.A., Myers S.T., 2000, IAU Symposium 201, 47
\bibitem[2000]{Bullock00} Bullock J.S., et al., 2000, preprint [astro-ph/9908159]
\bibitem[2001]{Bullock01} Bullock J.S., Dekel, A., Kolatt, T.S., Primack, J.R., \& 
  Somerville, R.S., 2001, ApJ, 550, 21
\bibitem[2001]{Cohn01}
  Cohn J.D., Kochanek C.S., McLeod B.A., Keeton C.R., 2001,
  ApJ, in press [astro-ph/0008390]
\bibitem[1994]{CAFNZ94}
  Cole S., et al., 1994, MNRAS, 271, 781
\bibitem[1989]{ColKai}
  Cole S., Kaiser N., 1989, MNRAS, 237, 1127
\bibitem[2000]{Cole00}
  Cole S., Lacey C.G., Baugh C.M., Frenk C.S., 2000, MNRAS, in press
  [astro-ph/0007281]
\bibitem[1997]{DalSpeSum}
  Dalcanton J.J., Spergel D.N., Summers F.J., 1997, ApJ, 482, 659
\bibitem[2001]{deBlok01}
  de Blok, W.J.G., McGaugh, S.S., Bosma, A., \& Rubin, V.C., 2001, astro-ph/0103102
\bibitem[1998]{Dehnen98}
  Dehnen, W., \& Binney, J., 1998, MNRAS, 294, 429
\bibitem[1988]{efstathiou98}
  Efstathiou, G., Ellis, G., \& Peterson, B.A., 1988, MNRAS, 232, 431
\bibitem[1976]{faber76}
  Faber, S.M., \& Jackson, R.E., 1976, 204, 668
\bibitem[1998]{falco98}
  Falco, E. E., Kochanek, C. S., \& Munoz, J.A., 1998, ApJ, 494, 47
\bibitem[1994]{Flores94}Flores, R., \& Primack, J.R., 1994, ApJ, 427, L1
\bibitem[1998]{Fuk98}
  Fukugita M., Hogan C.J., Peebles P.J.E., 1998, ApJ, 503, 518
\bibitem[1991]{FT91}
  Fukugita M., Turner E.L., 1991, MNRAS, 253, 99
\bibitem[2000]{GWBKP}
  Gonzalez A.H., et al., 2000, ApJ, 528, 145 [astro-ph/9908075]
\bibitem[1999]{helbig99}
  Helbig, P., Marlow, D., Quast, R., Wilkinson, P. N., Browne, I. W. A., \& Koopmans, 
  L. V. E. 1999, A\&AS, 136, 297
\bibitem[1990]{Her}
  Hernquist L., 1990, ApJ, 356, 359
\bibitem[2000]{JFWCCEY}
  Jenkins A., Frenk C.S., White S.D.M., Colberg J.M., Cole S., Evrard A.E.,
  Yoshida N., 2000, MNRAS, in press [astro-ph/0005260]
\bibitem[1999]{KCDW}
  Kauffmann G., Colberg J.M., Diaferio A., White S.D.M., 1999,
    MNRAS, 303, 188
\bibitem[1993]{KauWhiGui}
  Kauffmann G., White S.D.M., Guiderdoni B., 1993, MNRAS, 264, 201
\bibitem[2000]{keeton00}
  Keeton, C. R., Falco, E. E., Impey, C. D., Kochanek, C. S., Lehar, J.,
  McLeod, B. A., Rix, H.-W., Munoz, J. A., \& Peng, C. Y. 2000, ApJ, in press (astro-ph/0001500)
\bibitem[2000]{Kee00}
  Keeton C.R., Madau P., 2000, ApJ submitted [astro-ph/0101058]
\bibitem[1998]{Kee98b}
  Keeton C.R., Kochanek C.S., Falco E.E., 1998, ApJ, 509, 561
\bibitem[1998]{Kee98}
  Keeton C., 1998, Harvard University PhD thesis.
\bibitem[1996]{KitSut}
  Kitayama T., Suto Y., 1996, ApJ, 469, 480
\bibitem[1993]{kochanek93a}
  Kochanek, C. S. 1993a, ApJ, 419, 12  % statistics
\bibitem[1995]{Koc95}
  Kochanek C.S., 1995, ApJ, 453, 545
\bibitem[1996]{Koc96}
  Kochanek C.S., 1996, ApJ, 466, 638
\bibitem[2001a]{Koc00}
  Kochanek C.S., Pahre M.A., Falco E.E., Huchra J.P., Mader J.,
  Jarrett T.H., Chester T., Cutri R., Schneider S.E., 2001,
  ApJ in press [astro-ph/0011456]
\bibitem[2001]{kochanek01a}
  Kochanek, C.S., \& White, M., 2001, ApJ in press, [astro-ph/0102334]
\bibitem[2001b]{kochanek01b}
  Kochanek, C.S., Pahre, M.A., \& Falco, E.E., 2001, submitted to ApJ, [astro-ph/0011458]
\bibitem[1994]{LacCol}
  Lacey C., Cole S., 1994, MNRAS, 271, 676
\bibitem[1991]{LacSil}
  Lacey C., Silk J., 1991, ApJ, 381, 14
\bibitem[2000]{LiOst}
  Li L-X., Ostriker J.P., 2000, preprint [astro-ph/0010432]
\bibitem[2000]{Mahdavi00}
   Mahdavi, A., Bohringer, H., Geller, M.J., \& Ramella, M., 2000, ApJ, 534, 114
\bibitem[1993]{maoz93}
  Maoz, D., \& Rix, H.-W. 1993, ApJ, 416, 425
\bibitem[1997]{Maoz97}
  Maoz D., Rix H.-W., Gal-Yam A., Gould A., 1997, ApJ, 486, 75
\bibitem[1998]{MMW}
  Mo H.J., Mao S., White S.D.M., 1998, MNRAS, 295, 319
\bibitem[1994]{Moore94}Moore, B., 1994, Nature, 370, 629
\bibitem[1998]{MGQSL}
  Moore B., et al., 1998, ApJ, 499, L5
\bibitem[2000]{Mortlock00}
  Mortlock D.J., Webster R.L., 2000, MNRAS, 319, 872 [astro-ph/0008081]
\bibitem[1988]{Nar88}
  Narayan R., White S.D.M., 1988, MNRAS, 231, 97P
\bibitem[1996]{NFW}
  Navarro J., Frenk C.S., White S.D.M., 1996, ApJ, 462, 563
\bibitem[2000]{NewDav}
  Newman J.A., Davis M., 2000, ApJ, 543, L11 
\bibitem[2201]{Pahre01}
  Pahre, M.A, Kochanek, C.S., \& Falco, E.E., 2001, in preparation
\bibitem[2000]{Peacock00}
  Peacock, J.A. \& Smith, R.E., 2000, MNRAS, 318, 1144
\bibitem[1999]{Pearce99}
  Pearce F.R., Jenkins A., Frenk C.S., Colberg J.M., White S.D.M.,
  Thomas P.A., Couchman H.M.P., Peacock J.A., Efstathiou G., 1999,
  ApJ, 521, L99
\bibitem[2000]{Phil00}
  Phillips P.M., Browne I.W.A., Wilkinson P.N., Jackson N.J., 2000,
  IAU Symposium 201 [astro-ph/0011032]
\bibitem[2000]{PorMad}
  Porciani C., Madau P., 2000, ApJ, 532, 679
\bibitem[1974]{PreSch}
  Press W., Schechter P., 1974, ApJ, 187, 425
\bibitem[1994]{Rix94}
  Rix H.-W., Maoz D., Turner E.L., Fukugita M., 1994, ApJ, 435, 49
\bibitem[2001]{Salucci01}
  Salucci, P., 2001, MNRAS, 320, L1
\bibitem[2001]{Scoccimarro01}
   Scoccimarro, R., Sheth, R.K., Hui, L., \& Jain, B., 2001, ApJ, 546, 20 
\bibitem[1999]{SheTor}
  Sheth R., Tormen G., 1999, MNRAS, 308, 119 (1999)
\bibitem[1993]{Shim93}
  Shimasaku K., 1993, ApJ, 413, 59
\bibitem[2001]{SKBKKPD}
  Sigad Y., et al., preprint [astro-ph/0005323]
\bibitem[1997]{skrutskie97}
  Skrutskie, M. F., \etal\  1997, in The Impact of Large Scale Near-IR Sky Surveys,
  F. Garzon et al., eds., (Dordrecht: Kluwer) 187
\bibitem[1999]{SomPri}
  Somerville R., Primack J., 1999, MNRAS, 310, 1087
\bibitem[1977]{tully77}
  Tully, R. B., \& Fisher, B.  1977, A\&A, 54, 661
\bibitem[1984]{TOG84}
  Turner E.L., Ostriker J.P., Gott J.R., 1984, ApJ, 284, 1
\bibitem[2000]{vandenBosch00}
  van den Bosch, F.C., Robertson, B.E., Dalcanton, J.J., \& de Blok, W.J.G., 2000, AJ, 199, 1579
\bibitem[2001]{vandenBosch01}
  van den Bosch, F.C. \& Swaters, R.A., 2001, MNRAS, 325, 1017
\bibitem[1995]{Wamb95}
  Wambsganss J., Cen R., Ostriker J.P., Turner E.L., 1995,
  Science, 268, 274
\bibitem[1998]{Wamb98}
  Wambsganss J., Cen R., Ostriker J.P., 1998, ApJ, 494, 29
\bibitem[2001]{WhiHerSpr}
  White M., Hernquist L., Springel V., 2001, ApJL, 550, L129
\bibitem[1991]{WhiFre}
  White S.D.M., Frenk C.S., 1991, ApJ, 379, 52
\bibitem[2001]{}
  Winn, J. N., Hewitt, J. N., \& Schechter, P. L.  2001, in Gravitational Lensing: Recent 
  Progress, Future Goals, ASP Conf. Series, eds. T. Brainerd \& C. S. Kochanek 
  (San Francisco:  ASP) (astro-ph/9909335)
\bibitem[1999]{Wu99}
  Wu X.-P., Xue Y.-J., Fang L.-Z., 1999, ApJ, 524, 22
\bibitem[2000]{Wy00}
  Wyithe, J.S.B., Turner, E.L., Spergel D.N., 2000, ApJ, submitted
  [astro-ph/0007354]
\end{thebibliography}
\end{document}